\documentclass[rnaas, modern]{aastex63}

\shorttitle{White Dwarf Merger Candidate}
\shortauthors{Garnavich et al.}

\newcommand{\thestar}{J005311}
\newcommand{\kms}{km$\;$s$^{-1}$}

\begin{document}

\title{Rapid Variability in the Wind from the White Dwarf Merger Candidate J005311}

\author{Peter Garnavich}
\affiliation{Department of Physics, University of Notre Dame, Notre Dame, IN 46556}
\author{Colin Littlefield}
\affiliation{Department of Physics, University of Notre Dame, Notre Dame, IN 46556}
\affiliation{Department of Astronomy, University of Washington, Seattle, WA}
\author{Richard Pogge}
\affiliation{Department of Astronomy, The Ohio State University, Columbus, OH 43210}
\affiliation{Center for Cosmology \&\ AstroParticle Physics, The Ohio State University, Columbus, OH 43210}
\author{Charlotte Wood}
\affiliation{Department of Physics, University of Notre Dame, Notre Dame, IN 46556}

\correspondingauthor{Peter Garnavich}
\email{pgarnavi@.nd.edu}


\begin{abstract}

We analyze time-series spectroscopy of the white dwarf merger candidate \thestar\ and confirm the unique nature of its optical spectrum. We detect an additional broad emission feature peaking at 343~nm that was predicted in the \citet{gvaramadze20} models. Comparing ten spectra taken with the Large Binocular Telescope (LBT), we find significant variability in the profile of the strong OVI 381.1/383.4~nm emission feature. This appears to be caused by rapidly shifting subpeaks generated by clumpiness in the stellar wind of \thestar . This line variability is similar to what is seen in many Wolf-Rayet stars. However, in \thestar , the rate of motion of the subpeaks appears exceedingly high as they can reach 16000~\kms\ in less than two hours.

\end{abstract}

\section{Introduction}

\citet{gvaramadze20} recently identified a white dwarf merger (WDM) candidate, \thestar . Its optical spectrum is dominated by high ionization lines of oxygen similar to a WO-type Wolf-Rayet star \citep{tramper15}, but with no detectable helium emission. The maximum expansion velocity of the wind from \thestar\ is 16000~\kms , which is more than three times higher than seen in WO stars. Models suggest that \thestar\ is significantly more compact than massive WO stars and that its wind is driven by rapid rotation of a strong magnetic field \citep{gvaramadze20, kashiyama19}, thus supporting its identification as a WDM.

Many Wolf-Rayet winds are clumpy as shown by emission line profiles that vary on time scales of hours \citep{moffat2000}. Clumps are seen as subpeaks that appear to accelerate away from the line centers. Statistical properties of the winds can be derived by detailed studies of the subpeak motions. Here, we analyze a limited number of \thestar\ spectra to search for emission line profile variations in this unique object. 

\vfil

\section{Data}

We obtained spectra of \thestar\ using the Large Binocular Telescope (LBT\footnote{The LBT is an international collaboration among institutions in the United States, Italy and Germany. LBT Corporation partners are: The University of Arizona on behalf of the Arizona Board of Regents; Istituto Nazionale di Astrofisica, Italy; LBT Beteiligungsgesellschaft, Germany, representing the Max-Planck Society, The Leibniz Institute for Astrophysics Potsdam, and Heidelberg University; The Ohio State University, representing OSU, University of Notre Dame, University of Minnesota and University of Virginia.}) and twin Multi-Object Dual Spectrographs \citep[MODS;][]{pogge12} on 2020, September 11 (UT). Five, 540s exposures were taken with MODS1 through the SX telescope providing an image cadence of 11~minutes with readout and overheads. The same exposure sequence was obtained with MODS2 through the DX telescope, but the UT time for these images averaged one minute later than for MODS1.

\begin{figure}
    \centering
    \includegraphics[width=\textwidth]{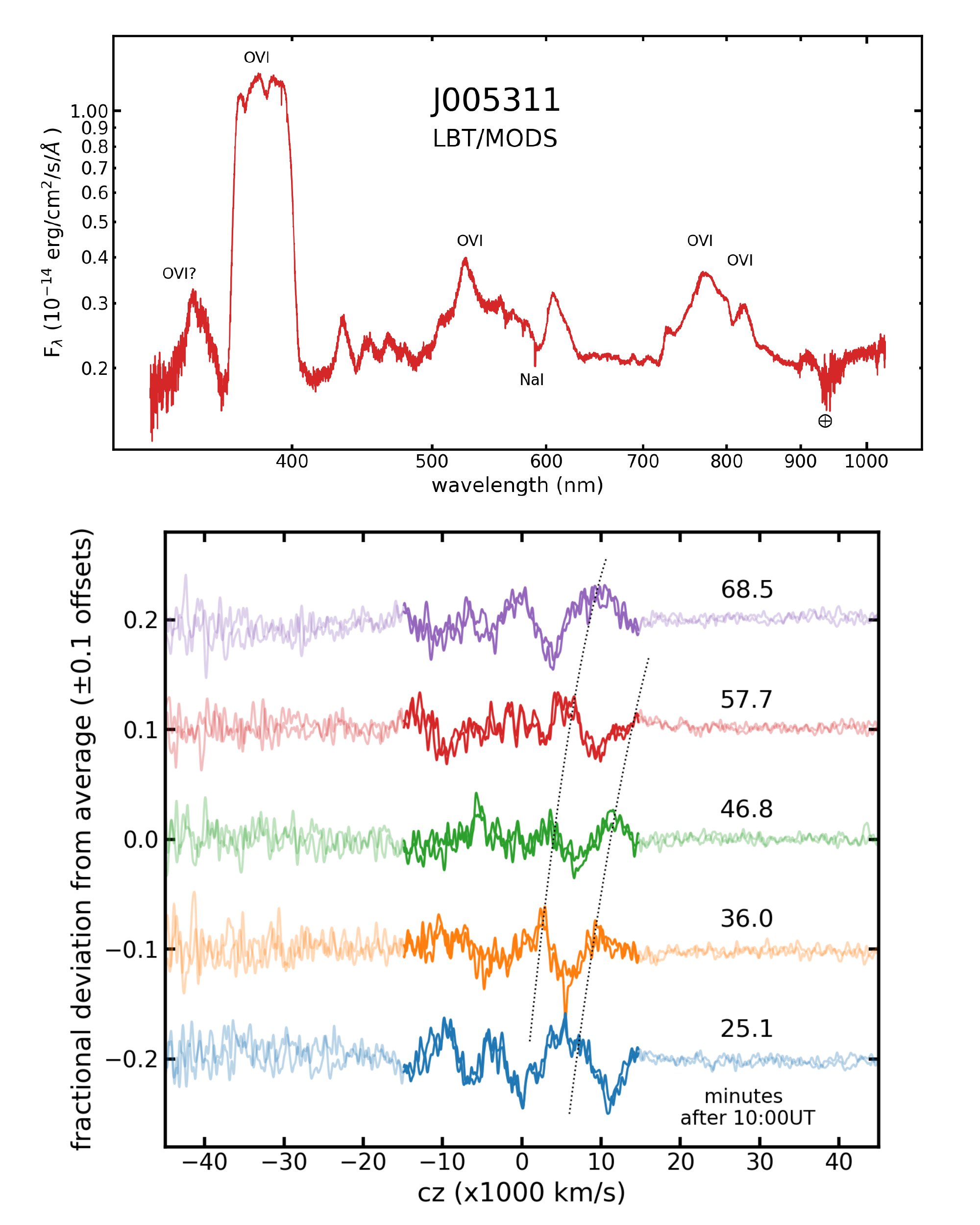}
    \caption{{\bf top:} The average LBT/MODS spectrum of \thestar\ with several of the OVI emission features identified \citep{gvaramadze20}. Narrow interstellar/circumstellar absorption lines due to NaI and CaII (H+K) are also visible.    {\bf bottom:} Changes in the OVI 381.1/383.4~nm emission profile over the ten individual exposures from the two telescopes. Each major epoch is shown shifted in the y-axis by a value of 0.1. The times for exposures from MODS1 and MODS2 differ by one minute and are plotted together. The mean UT of their exposures is shown at the right. The plausible motion of two wind subpeaks is indicated by the dotted lines.   }
    \label{fig1}
\end{figure}

One-dimensional spectra were extracted from the CCD images using reduction packages in IRAF, wavelength calibrated with neon and argon emission line lamps, and flux calibrated using the spectrophotometric standard star G191b2b obtained on the same night. Major telluric absorption features in the red were removed by dividing the \thestar\ by a normalized hot white dwarf spectrum. Smoke from major fires in California significantly dimmed \thestar\ during these observations, so the overall flux was corrected to match the expected brightness of $V=15.4$ mag. The resulting average spectrum is displayed in the top panel of Figure~\ref{fig1}.

\section{Analysis}

Overall, our spectrum is consistent with that obtained by \citet{gvaramadze20}. The OVI emission lines, including the exceptional 381.1/383.4~nm doublet are evident. A strong emission feature at 343~nm, predicted by \citet{gvaramadze20}, is detected at the ultraviolet end of our data and it is likely also OVI.  

We search for variability in the 381.1/383.4~nm OVI emission by normalizing and averaging the five spectra obtained on MODS1 and subtracting the average from the individual spectra. The resulting profile variations are shown in the lower panel of Figure~\ref{fig1}. The line width is $\pm 16000$~\kms , so only the residuals within that velocity range are highlighted. The spectra are offset in the vertical direction to indicate time increasing from bottom to top. Variable subpeaks and troughs are detected with an amplitude of as much as 4\%\ of the average flux.

The analysis was repeated for spectra obtained by MODS2 and the results over-plotted on the MODS1 emission line residuals. The exposures from the two telescopes overlap by 80\%\ in time. The consistency in line variations between these independent data sets confirms that we are seeing true fluctuations in the line profile.

The substructure must be changing rapidly to avoid being averaged out. We mark a plausible connection between subpeaks on the redshifted side of the 381.1/383.4~nm line. If this does represent the motion of long-lived subpeaks, then the rate of velocity change is approximately 200~\kms~minute$^{-1}$. The spectral resolution at the bright OVI emission is 170~\kms\ so the doublet lines would be resolved if the subpeaks were narrow. Instead, the clumps widths are typically 2000~\kms , and this is likely a combination of the the rapid acceleration over our 9-minute integrations and the blending of the doublet lines within a subpeak.

Simulations of these data assuming 2000~\kms -wide (FWHM) subpeaks shifting at 200~\kms~minute$^{-1}$ generally reproduce the observed redshifted side of the emission residuals. Because of the limited number of exposures used in building an average line profile, the simulations suggest that the true subpeak amplitudes are 20\%\ larger than they appear in the data. Subpeaks shifting as slowly as 50~\kms~minute$^{-1}$ remain detectable, but simulations show them as narrow peak/trough pairs. These slowly shifting subpeaks may be the cause of the narrow structures visible around zero velocity.  Overall, a faster cadence and longer time sequence is needed to fully characterize the variable emission line structures in \thestar .

\section{Conclusion}\label{sec:discussion}

LBT/MODS spectra of the WDM candidate \thestar\ confirms the unique nature of its optical spectrum. We detect an additional broad emission feature peaking at 343~nm that was predicted in the \citet{gvaramadze20} models. We find significant variability in the profile of the strong OVI 381.1/383.4~nm emission feature. This appears to be caused by rapidly shifting subpeaks generated by clumpiness in the stellar wind of \thestar\ and is similar to what is seen in many Wolf-Rayet stars. However, in \thestar , the rate of motion of the subpeaks appears exceedingly high as they can reach the 16000~\kms\ edge of the emission line in less than two hours.

\acknowledgments

We thank O. Kuhn for help in obtaining the LBT/MODS spectra. This paper used data obtained with the MODS spectrographs built with funding from NSF grant AST-9987045 and the NSF Telescope System Instrumentation Program (TSIP), with additional funds from the Ohio Board of Regents and the Ohio State University Office of Research.

{}

\end{document}